\documentclass{article}
\usepackage{spconf,amsmath,graphicx,hyperref}
\usepackage{url}
\usepackage{cite}
\usepackage{amsmath,amssymb,amsfonts}
\usepackage{algorithmic}
\usepackage{graphicx}
\usepackage{textcomp}
\usepackage{xcolor}
\usepackage{float}
\usepackage{subfig}
\usepackage{hyperref}
\usepackage{enumitem}
\usepackage{booktabs} 
\usepackage{siunitx}
\usepackage{graphicx}

\title{Emotional Dimension Control in Language Model-Based Text-to-Speech: Spanning a Broad Spectrum of Human Emotions}
\name{Kun Zhou$^{1}$, You Zhang$^{2}$, Dianwen Ng$^{1}$, Shengkui Zhao$^{1}$, Hao Wang$^{1}$, Bin Ma$^{1}$}
\address{$^{1}$Tongyi Lab, Alibaba Group, Singapore $^{2}$University of Rochester, United States of America}
\begin{document}
\ninept
\maketitle
\begin{abstract}
Emotional text-to-speech (TTS) systems struggle to capture the full spectrum of human emotions due to the inherent complexity of emotional expressions and the limited coverage of existing emotion labels. To address this, we propose a language model-based TTS framework that synthesizes speech across a broad range of emotional styles. Our approach enables flexible user control along three continuous dimensions—pleasure, arousal, and dominance (PAD). To enable this, we train an emotional dimension predictor that maps categorical emotion labels in speech datasets into the PAD space, grounded in established psychological research. Importantly, while the emotional dimension predictor leverages categorical labels, the TTS framework itself does not require explicit emotion labels during training. Objective and subjective evaluations demonstrate that our framework effectively generates more expressive emotional styles and enhances both naturalness and diversity compared to baselines.
\end{abstract}
\section{Introduction}
\label{sec:intro}
Emotional text-to-speech (TTS) aims to generate speech from text that conveys human-like emotions \cite{triantafyllopoulos2023overview}. Recent advances in deep learning have significantly enhanced its capabilities. As a pivotal technology for human-computer interaction, emotional TTS facilitates the development of dialogue systems that are both engaging and empathetic \cite{triantafyllopoulos2024expressivity}.

However, current emotional TTS systems remain constrained by both the inherent complexity of human emotions and the narrow set of labels available in existing speech databases \cite{triantafyllopoulos2024expressivity}. As a result, they can only generate a limited pre-defined set of emotions. Psychological research suggests that humans can experience around 34,000 distinct emotions \cite{plutchik2001nature}, far beyond the expressive range captured by current systems. Prior research in emotional TTS has largely followed two paradigms: (1) supervised learning with categorical emotion labels \cite{zhou2022emotion,yoon2022language}; and (2) style transfer using reference speech \cite{stanton2018predicting}. Both approaches, however, suffer from the limited diversity of emotional expressions present in existing emotional speech datasets, often leading to synthesized speech that sounds average, stereotypical, and lacking in fine-grained emotional control. More recently, researchers have explored hierarchical emotion distributions for fine-grained editing and text-driven control \cite{inoue2024hierarchical,inoue2024fine}. While these methods enhance fine-grained expressiveness, they remain fundamentally dependent on labeled emotion data, leaving the broader spectrum of human emotional experience underrepresented.

Recent language model (LM)-based TTS frameworks \cite{wang2023neural,zhang2023speak} show strong in-context learning capabilities. With sufficient training data, these models effectively render appropriate prosody for complex text inputs and exhibit robust zero-shot speaker cloning capabilities. Unlike traditional TTS tasks, which focus primarily on capturing a wide range of phonetic variations across speakers \cite{zhou2024phonetic}, emotional TTS emphasizes rendering diverse emotional styles, thus requiring large and varied emotional speech datasets with detailed emotion annotations. Emotional TTS systems like CosyVoice \cite{du2024cosyvoice} and EmoCtrl-TTS \cite{wu2024laugh}, built on state-of-the-art language models, often training with thousands of hours of speech data, highlighting the cost of data collection. However, publicly available emotional speech datasets are typically small and lack of detailed annotations\cite{zhou2021emotional}, making it difficult for training a robust LM-based TTS system from scratch. A common strategy is to fine-tune a pre-trained TTS model using small emotional datasets \cite{zhou2021limited}, but this approach risks overfitting and limits generalization across emotions and speakers \cite{zhou2020converting}. To overcome these limitations, we propose a framework that controls speech synthesis through continuous emotional dimensions, enabling the generation of nuanced emotions beyond those present in training data. Leveraging LMs’ in-context learning, our approach enables flexible emotion rendering and broadens the set of emotions the system can synthesize, thereby expanding TTS’s expressive range even with limited or unlabeled emotional data.

In this paper, we propose an effective approach for LM-based TTS systems to generate diverse emotional styles by controlling an emotional dimension vector. Our method draws inspiration from Russell's emotion theory \cite{russell1977evidence}, which identifies three independent dimensions—pleasure, arousal, and dominance—as sufficient to capture the complexity of emotional expressions across various contexts and individuals. Using these dimensions as foundational anchors, we develop a model that predicts their values. The pre-trained emotion dimension  (ED) predictor is then integrated into an LM-based TTS system to assist in next-token prediction, alongside the speaker embedding. During inference, our framework enables direct control over emotional dimensions, allowing the synthesis of diverse emotional styles without requiring emotional reference inputs. Our contributions are summarized as follows:

\setlength{\leftmargini}{15pt} 
\begin{itemize}
\setlength{\itemsep}{5pt}      
\setlength{\parskip}{-1pt}      
\setlength{\itemindent}{0pt}   
\setlength{\labelsep}{5pt}
    \item We propose a novel LM-based TTS framework that controls emotions through continuous emotional dimensions, moving beyond the limitations of conventional categorical labels and costly dimensional annotations. Unlike prior approaches, our model learns expressive emotional styles directly from speech, without requiring explicit emotion labels during TTS training;
    \item We introduce a psychologically grounded control space, where users can directly manipulate a low-dimensional emotional vector (Pleasure–Arousal–Dominance) to explore and synthesize a wide spectrum of emotions with fine granularity, extending beyond those predefined emotions observed in training data;
    \item Experiments show that our approach not only improves the naturalness of zero-shot emotion cloning but also consistently outperforms a strong LM-based TTS baseline in terms of emotional intelligibility with subjective evaluation.
\end{itemize}
We introduce our proposed method in Section 2. In Section 3, we report our experimental results. Section 4 concludes the paper.

\begin{figure*}
    \centering
    \includegraphics[width=0.9\linewidth]{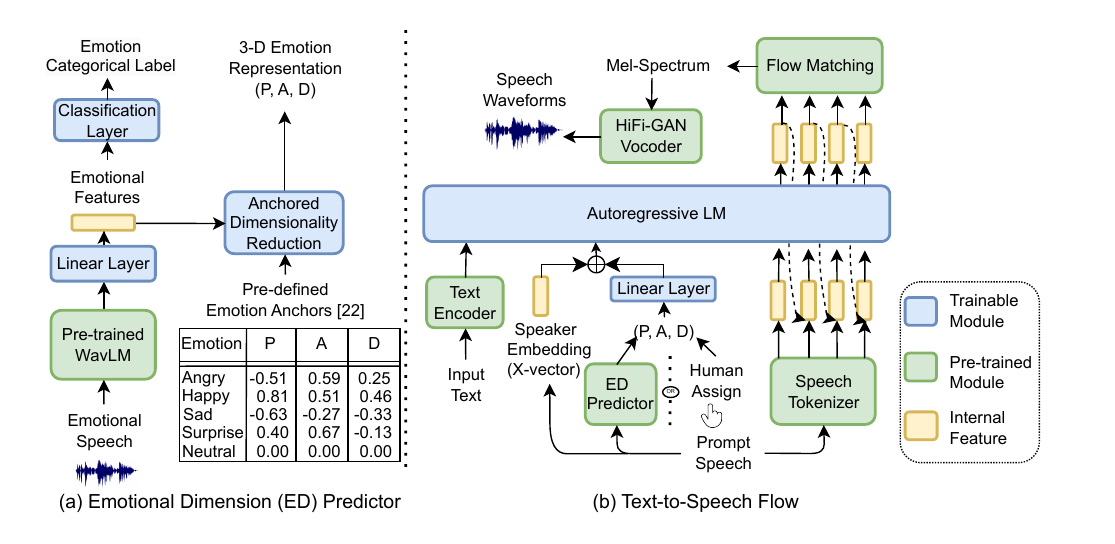}
   \vspace{-5mm}
    \caption{An overview of the proposed text-to-speech (TTS) framework with emotional dimension control, consisting of: (a) Emotional 
    Dimension (ED) Predictor Training, and (b) Text-to-Speech Flow. The ED predictor is pre-trained on an emotional speech dataset to map emotional features to dimension representations via anchored dimensionality reduction. It then guides the autoregressive language model (LM) to predict acoustic details. `P', `A', and `D' denote `Pleasure', `Arousal', and `Dominance'. PAD values can be either inferred from prompt speech (`Emotion Cloning') or assigned by humans (`Emotion Control').
    }
    \label{fig:main}
\end{figure*}

\section{Proposed Method}

\subsection{Problem Formulation}
We formulate TTS as a conditional language modeling task, where the speech waveform is quantized into discrete tokens and is then modeled with a language model \cite{wang2023neural,du2024cosyvoice,zhou2024phonetic}.
Since training language models typically requires large-scale data to achieve robust performance \cite{kim2024clam,lajszczak2024base}, this is often impractical for emotional TTS, as emotional speech data are scarce and costly to collect. To address this, we represent speech emotion as continuous dimensions of expressiveness that can be learned directly from expressive speech data \cite{lotfian2017building}. Our goal is to develop a TTS model that learns from such data during training and synthesizes speech that conveys intended emotions.

Our TTS system adopts an autoregressive language model (LM) decoder, a flow-matching module, and a HiFi-GAN vocoder. The LM decoder first generates discrete acoustic tokens from text, which the flow-matching module converts into Mel-spectrograms. Finally, HiFi-GAN synthesizes the waveform from the predicted Mel-spectrograms. Beyond this baseline design, our model introduces two key enhancements: (1) emotional dimension (ED) embeddings for fine-grained emotion control, and (2) conditioning on a reference speech to improve voice consistency.


\subsection{Predicting Emotional Dimensions from Categories}
Human emotions can be represented categorically \cite{ekman1992argument} or dimensionally \cite{russell1980circumplex}. Compared to the categorical emotional approach~\cite{ekman1992argument} that lacks the capability to capture subtle emotional differences, the dimensional approach, such as Russell’s circumplex model, views emotions as fluid experiences across several dimensions, offering a more nuanced understanding \cite{russell1980circumplex}. 
The specific definitions \cite{russell1977evidence} of 151 emotion types, as rated by 300 subjects, were reliably measured on a pleasure-arousal-dominance scale as shown in Table \ref{tab:pad}.

Despite these advances, most emotional TTS research still relies on a limited set of categorical emotion labels~\cite{barra2010analysis,zhou2020converting}. Only a few studies have explored generating mixed emotional styles by combining basic emotions~\cite{zhou2022speech,tang2023emomix}. Recent TTS systems that incorporate emotional dimensions~\cite{cho2024emosphere,kalyanemotion,wu2024laugh} typically depend on SER models trained with costly and labor-intensive dimensional annotations. As a result, they inherit well-known SER biases—such as subjective and inconsistent labeling and imbalanced emotion distributions—which limit their ability to generalize when applied to generation tasks~\cite{wagner2023dawn,parthasarathy2017jointly}.

Inspired by a recent SER study~\cite{zhou2024learning} predicting arousal-valence only from categorical emotions, 
we design an ED predictor to map discrete emotions into a pleasure-dominance-arousal space (Figure~\ref{fig:main}(a)). We train the ED predictor with an emotional speech dataset, and each emotion category is assigned an \textit{anchor} based on \cite{russell1977evidence} to guide training, with initial vectors perturbed by a Gaussian noise $\theta$ = 0.01. Using pre-trained WavLM \cite{chen2022wavlm} for feature extraction, we predict emotional labels through a linear and classification layer.   We derive a 128-d emotional feature vector from the linear layer and fine-tune the ED vectors from their initial settings. We create a $k$-Nearest Neighbor (kNN) graph and integrate categorical labels to differentiate emotion classes. UMAP \cite{mcinnes2020} refines the vectors by minimizing cross-entropy loss between high-dimensional features and ED vectors. The ED predictor then estimates pleasure-arousal-dominance values for new speech samples, guiding speefch token predictions in the TTS model.

\begin{table}[t]
\centering \scriptsize
    \tabcolsep=0.12cm
    \renewcommand{\arraystretch}{1}
    
\caption{Examples of emotional state definitions in terms of the three dimensions of Pleasure (`P'), Arousal (`A'), and Dominance (`D'), following \cite{russell1977evidence}.}

\scalebox{1}{

\begin{tabular}{l|S[table-format=-1.2] S[table-format=-1.2] S[table-format=-1.2] || l| S[table-format=-1.2] S[table-format=-1.2] S[table-format=-1.2]}
\hline
\multicolumn{1}{c|}{\textbf{Emotion}} & \textbf{P} & \textbf{A} & \textbf{D} &
\multicolumn{1}{c|}{\textbf{Emotion}} & \textbf{P} & \textbf{A} & \textbf{D} \\
\hline
Angry    & -0.51 &  0.59 &  0.25 & Excited   &  0.62 &  0.75 &  0.38 \\
Happy    &  0.81 &  0.51 &  0.46 & Alert     &  0.49 &  0.57 &  0.45 \\
Sad      & -0.63 & -0.27 & -0.33 & Protected &  0.60 & -0.22 & -0.40 \\
Surprise &  0.40 &  0.67 & -0.13 & Relaxed   &  0.68 & -0.46 &  0.20 \\
Anxious  &  0.01 &  0.59 & -0.15 & Neutral   &  0.00 &  0.00 &  0.00 \\
\hline
\end{tabular}}
\vspace{-5mm}
\label{tab:pad}
\end{table}

\subsection{TTS Training with Emotional Dimension Guidance}
The overall TTS diagram is illustrated in Figure \ref{fig:main}(b). Given prompt speech,
emotional dimension vectors are extracted using the pre-trained ED predictor (Figure \ref{fig:main}(a)), while the speaker embedding is obtained from a pre-trained voiceprint model\footnote{\url{https://github.com/alibaba-damo-academy/3D-Speaker/tree/main/egs/3dspeaker/sv-cam++}}. The input text is first processed through a Grapheme-to-Phoneme tokenizer and then mapped to a shared semantic space with speech tokens using a text encoder. These extracted features are concatenated and serve as conditioning inputs for the speech tokenizer, which produces speech token sequences.

We then frame the TTS task as an auto-regressive speech token generation task using a language model (LM). During LM training, we employ a teacher-forcing scheme to optimize LM predictions, where the left-shifted sequence serves as the input, and the original sequence acts as the expected output. We compute the cross-entropy loss between the predicted and ground-truth speech tokens, ensuring effective learning of token dependencies and accurate reconstruction of speech sequences.

\subsection{Emotion Rendering with Emotional Dimension Control}
During inference, the emotional dimension (ED) vector can either be inferred from the prompt speech (`emotion cloning') or directly specified by the users (`emotion control'), allowing fine-grained adjustment of emotional expression. Given the input text and prompt speech, the language model autoregressively generates a sequence of speech tokens that encode both linguistic and prosodic information. These tokens are then passed to a pre-trained optimal-transport flow-matching model, which maps them to the distribution of Mel spectrograms. The flow-matching model produces high-fidelity spectrograms with smooth and natural prosody, which are subsequently converted to waveforms by a vocoder. In this way, our system synthesizes expressive and natural speech while preserving the desired emotional characteristics, whether inferred from the prompt or explicitly set by the user.

\begin{figure}[t]
    \centering
    \vspace{-10mm}
    \includegraphics[width=1\linewidth]{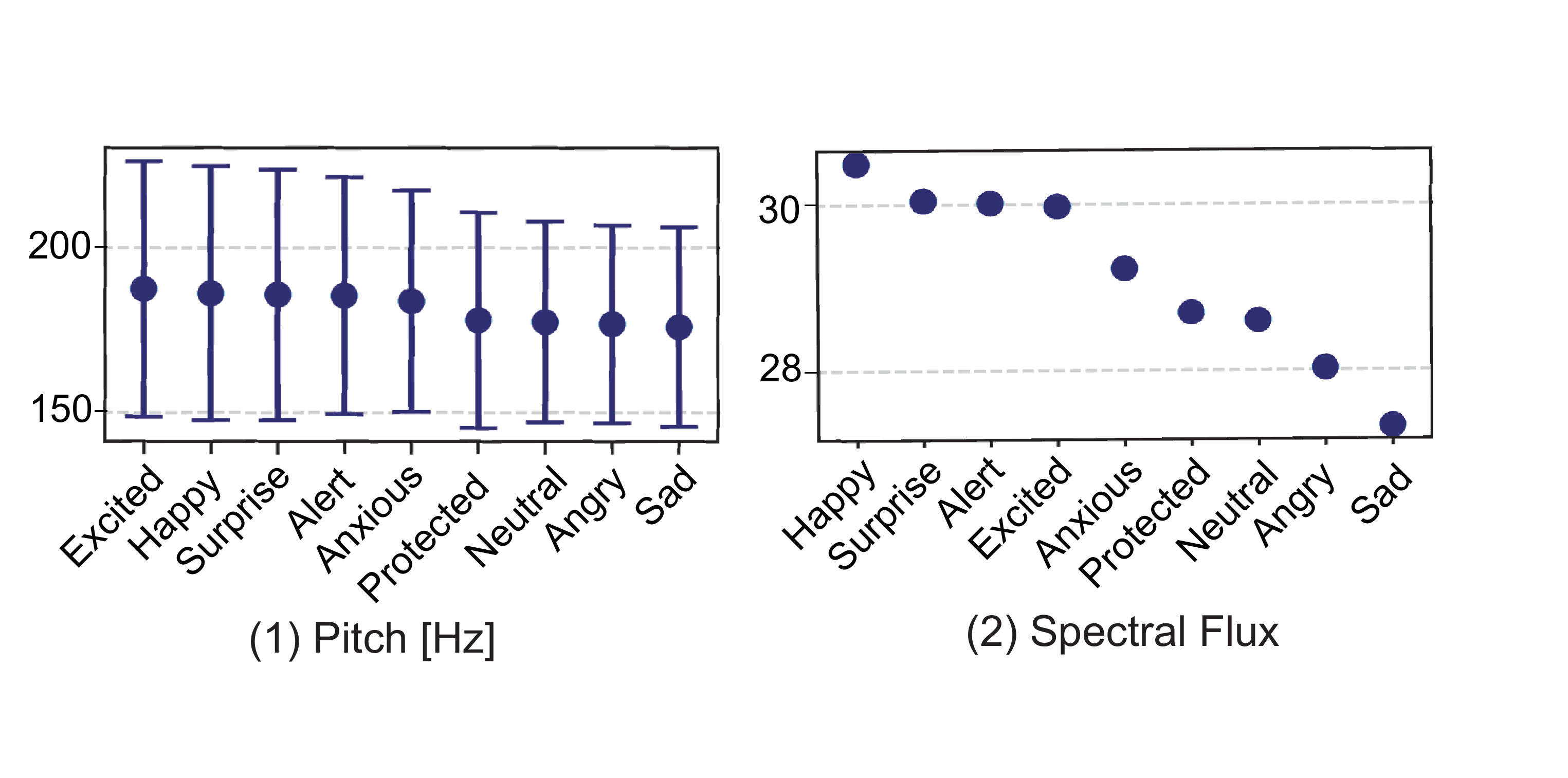}
    \vspace{-12mm}
    \caption{Statistical analysis of pitch and spectral flux for 9 emotions synthesized via ED Control in the proposed framework.}
    \label{fig:statics}
\end{figure}
\section{Experiments}

\subsection{Experimental Setup}
The emotional dimension predictor is trained on the training set from the English subset of the ESD dataset \cite{zhou2021seen}, which includes around 10 hours of emotional speech data across 5 basic emotions: neutral, angry, happy, sad, and surprise. The emotional dimension predictor reported an 81\% emotion classification accuracy on the ESD test set, which is comparable to reported SER baselines on this dataset \cite{zhou2021emotional}.
We train the TTS framework on the LibriTTS dataset \cite{Zen2019}, which consists of around 600 hours of speech data with transcriptions at a 24kHz sampling rate from 2,456 speakers. We merge the `train-clean' and `train-other' subsets as the training set and evaluate TTS performance on the `test-clean' subset. 
Note that while LibriTTS only contains expressive speech, it does not include any explicit emotion labels. 

For the experimental setup, we follow a prompt-based synthesis protocol. For each speaker in the LibriTTS test set, we randomly select one utterance from the same speaker as the acoustic prompt, ensuring it is distinct from the target content. The system then synthesizes speech for the test transcriptions conditioned on this prompt.

We evaluate our framework under two inference settings, which differ in how the emotional dimension (ED) values are obtained:
\begin{itemize}
    \item \textbf{Emotion Cloning}: ED values are automatically inferred from the prompt audio, enabling zero-shot emotion cloning;
    \item \textbf{Emotion Control}: ED values are manually specified based on psychological theory \cite{russell1977evidence} to synthesize speech corresponding to different emotions, as shown in Table \ref{tab:pad};
\end{itemize}
The emotion cloning setting best demonstrates the model’s ability to generalize to unseen emotions and speakers in a zero-shot manner. The emotion control setting highlights the flexibility of the framework by allowing users to explore and manipulate emotions beyond those present in training data.
The duration of each synthesized speech sample ranges from 3 to 10 seconds. Speech demos are publicly available online\footnote{Speech Demos: \url{https://demos46.github.io/emotion_pad/}}. 
\begin{table}[t]
\caption{Mean Opinion Score (`MOS') of naturalness for the zero-shot emotion cloning task.}
\centering\scalebox{0.85}{
\begin{tabular}{lll}
\hline
Ground Truth & \textbf{Emotion Cloning (Proposed)} & Baseline CosyVoice \cite{du2024cosyvoice} \\ \hline
 4.80 {$\pm$ 0.08}  & \textbf{4.54 {$\pm$ 0.18}} & 4.36 {$\pm$ 0.13}\\ \hline
\end{tabular}}
\label{tab:cmos}
\vspace{-4mm}
\end{table}
\subsection{Implementation Details}
The backbone TTS framework is similar to that of CosyVoice \cite{du2024cosyvoice}, which consists of a speech tokenizer, a text encoder, a text-to-token LM, and a flow-matching module. The speech tokenizer employs an ESPNet Conformer ASR model along with a vector quantizer of a codebook size of 4096 after the first 6 encoder layers. The text encoder consists of 6 transform layers with an 8-head and 512-d attention layer. The LM contains 12 transform layers with an 8-head and 512-d attention layer. An optimal-transport conditional flow matching model (OT-CFM) \cite{tongimproving} is employed to predict the Mel spectrograms. A HiFi-GAN\footnote{\url{https://github.com/jik876/hifi-gan}} vocoder recovers the speech signal from the predicted Mel spectrogram.
The emotional dimension predictor consists of a WavLM features extractor followed by a linear layer, and an anchored dimensionality reduction module. The WavLM model is fine-tuned for 100 epochs with a batch size of 64 using the Adam optimizer at a 0.0001 learning rate. Dimensionality reduction settings are a 0.1 minimum distance and 20 nearest neighbors, with a 0.01 learning rate for optimization.

\subsection{Objective Evaluation}
Pitch and spectral flux are well-established acoustic correlates of emotional expressiveness: higher pitch is often linked to heightened arousal in emotions such as `Excitement', while greater spectral flux, reflecting rapid spectral variation, is associated with dynamic emotions like `Surprise'. To assess whether our system reproduces these patterns, we compute pitch and spectral flux statistics for the synthesized speech, as shown in Figure \ref{fig:statics}.

The results reveal that emotions such as `Excitement', `Surprise', `Happy', and `Alert' exhibit relatively higher values for both features, suggesting greater vocal dynamism and expressiveness. In contrast, `Sad', `Angry', `Neutral', and `Protected' show lower values, indicating more subdued or restrained speech patterns. These trends are consistent with findings in established emotion theory \cite{russell1977evidence}, where similar acoustic distinctions across emotions have been reported. This alignment provides evidence that our model not only generates distinguishable emotional styles but also reflects genuine acoustic correlates of human emotional expression.

\begin{figure}[t]
    \centering
    \vspace{-5mm}
    \includegraphics[width=0.9\linewidth]{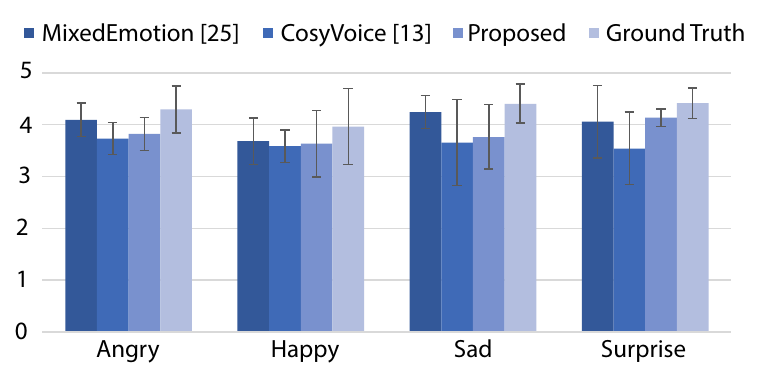}
    \caption{Mean Opinion Score for assessing emotional intelligibility (`E-MOS') of our proposed system for the emotion cloning task.}
    \vspace{-4mm}
    \label{fig:emos}
\end{figure}

\subsection{Subjective Evaluation}
We conducted listening experiments with 10 native English speakers, each evaluating a total of 260 synthesized speech samples. Such participant numbers and sample sizes are standard practice in TTS research \cite{triantafyllopoulos2024expressivity}. We first evaluate our proposed framework on the zero-shot emotion cloning task, where the ED is inferred from the prompt speech (`Proposed'). 
We selected CosyVoice \cite{du2024cosyvoice} as the baseline and trained it from scratch on LibriTTS to ensure a fair comparison with the proposed model.
A mean opinion score (MOS) test is conducted to evaluate the speech naturalness, where participants were asked to score each sample on a 5-point scale (5: Excellent, 4: Good, 3: Satisfactory, 2: Unsatisfactory, 1: Bad). As shown in Table \ref{tab:cmos}, our proposed framework achieves a score of 4.54, outperforming the baseline score of 4.36, while lower than the ground-truth score of 4.80. It shows that our proposed framework achieves better naturalness than the baseline CosyVoice.

We additionally conducted a mean opinion score test for emotional intelligibility (E-MOS), where participants rated each synthesized sample based on perceived emotion. Our framework was compared against two baselines: MixedEmotion \cite{zhou2022speech} and CosyVoice \cite{du2024cosyvoice}. As shown in Fig.~\ref{fig:emos}, our framework consistently achieves higher E-MOS scores than CosyVoice. MixedEmotion, which is trained on emotional data, attains the best score among systems but still falls below ground-truth speech. In contrast, both CosyVoice and our framework operate in a zero-shot setting, where no explicit emotional labels or emotional data are used during TTS training, and emotion rendering relies solely on conditioning with a reference speech prompt. Under this more challenging scenario, our method demonstrates advantages in emotional intelligibility.
These results indicate that our proposed method effectively captures and replicates emotional nuances in synthesized speech, thereby enhancing emotion cloning performance.

\begin{figure}[t]
    \centering
    \includegraphics[width=1\linewidth]{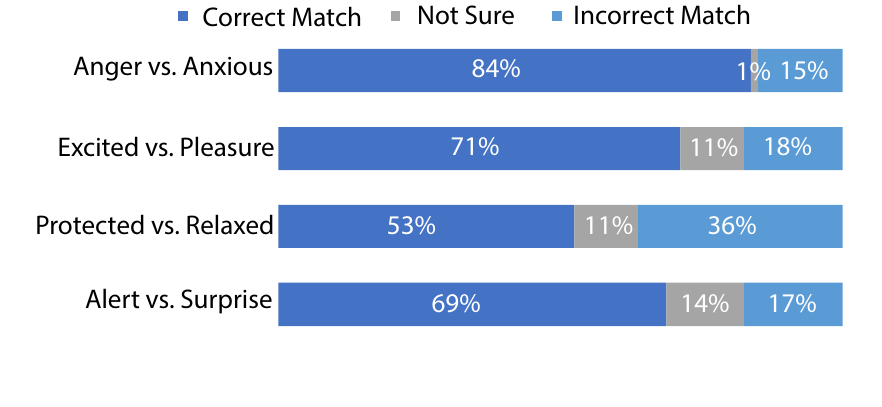}
    \vspace{-11mm}
    \caption{XAB test results for four synthesized emotion pairs, used to evaluate the intelligibility of perceived emotions and the effectiveness of emotion control in the proposed system.} 
    \vspace{-5mm}
    \label{fig:exp4}
\end{figure}

To further assess whether our system can synthesize distinguishable emotional categories, especially those with overlapping acoustic characteristics, we conducted XAB tests on four emotion pairs$^2$: 1) `Anger' vs. `Anxious'; 2) `Excited'  vs. `Pleasure'; 3) `Protected'  vs. `Relaxed' ; 4) `Alert'  vs.
`Surprise'. These pairs were selected because they are theoretically differentiated by specific PAD dimensions (e.g., dominance in Anger vs. Anxious), making them more challenging to disambiguate. For each pair, participants were asked to correctly match the emotions, such as choosing between option A: X is Anger and Y is Anxiety and option B: X is Anxiety and Y is Anger. The results are summarized in Fig. \ref{fig:exp4}. For all 4 emotion pairs, the correct match consistently exceeds 50\%, with the `Anger vs. Anxious' pair achieving the highest accuracy at around 84\%. 
The results in Fig.\ref{fig:exp4} demonstrate a positive alignment with 2 key findings in \cite{russell1977evidence}:
\setlength{\parskip}{-1pt} 
\begin{itemize}
    \item `Dominance' serves as a key factor in differentiating emotions, for example distinguishing `Angry' from `Anxious', `Alert' from `Surprised', and `Relaxed' from `Protected'\footnote{The first emotion in each pair involves dominance; the second emotion involves submissiveness.};
    \item The combination of elevated `Pleasure' and `Arousal' dimensions is commonly associated with `Excitement';
\end{itemize}
These results demonstrate that our TTS system can render emotions consistent with psychological theories of the multidimensional nature of affect, incorporating pleasure, arousal, and dominance. Unlike prior systems that rely on limited categorical labels or costly dimensional annotations and thus struggle to capture the full range of emotions, our model leverages continuous dimensions to synthesize speech with diverse emotional styles. This enhances both naturalness and emotional intelligibility, effectively bridging psychological theory with practical speech synthesis.
\section{Conclusion}
In this paper, we proposed an emotional TTS framework that integrates established emotion theory with language model-based speech generation. By controlling pleasure, arousal, and dominance, our proposed system synthesizes a wide range of emotional speech  without relying on large-scale emotion-labeled data, while preserving naturalness in zero-shot emotion cloning scenarios. This work highlights the value of combining psychological insights with data-driven modeling. Future directions include exploring dynamic within-utterance emotion control and multilingual adaptation, paving the way for more versatile, affect-aware TTS solutions.

\footnotesize
\bibliographystyle{IEEEbib}
\bibliography{mybib}

@article{triantafyllopoulos2023overview,
  title={An overview of affective speech synthesis and conversion in the deep learning era},
  author={Triantafyllopoulos, Andreas and Schuller, Bj{\"o}rn W and {\.I}ymen, G{\"o}k{\c{c}}e and Sezgin, Metin and He, Xiangheng and Yang, Zijiang and Tzirakis, Panagiotis and Liu, Shuo and Mertes, Silvan and Andr{\'e}, Elisabeth and others},
  journal={Proceedings of the IEEE},
  year={2023},
  publisher={IEEE}
}

@article{tongimproving,
  title={Improving and generalizing flow-based generative models with minibatch optimal transport},
  author={Tong, Alexander and FATRAS, Kilian and Malkin, Nikolay and Huguet, Guillaume and Zhang, Yanlei and Rector-Brooks, Jarrid and Wolf, Guy and Bengio, Yoshua},
  journal={Transactions on Machine Learning Research}
}

@inproceedings{Zen2019,
  author={Heiga Zen and Viet Dang and Rob Clark and Yu Zhang and Ron J. Weiss and Ye Jia and Zhifeng Chen and Yonghui Wu},
  title={{LibriTTS: A Corpus Derived from LibriSpeech for Text-to-Speech}},
  year=2019,
  booktitle={Proc. INTERSPEECH},
  pages={1526--1530},
  doi={10.21437/Interspeech.2019-2441},
}

@article{chen2022wavlm,
  title={Wavlm: Large-scale self-supervised pre-training for full stack speech processing},
  author={Chen, Sanyuan and Wang, Chengyi and Chen, Zhengyang and Wu, Yu and Liu, Shujie and Chen, Zhuo and Li, Jinyu and Kanda, Naoyuki and Yoshioka, Takuya and Xiao, Xiong and others},
  journal={IEEE Journal of Selected Topics in Signal Processing},
  volume={16},
  number={6},
  pages={1505--1518},
  year={2022},
  publisher={IEEE}
}

@article{wang2023neural,
  title={Neural codec language models are zero-shot text to speech synthesizers},
  author={Wang, Chengyi and Chen, Sanyuan and Wu, Yu and Zhang, Ziqiang and Zhou, Long and Liu, Shujie and Chen, Zhuo and Liu, Yanqing and Wang, Huaming and Li, Jinyu and others},
  journal={arXiv preprint arXiv:2301.02111},
  year={2023}
}

@article{yoon2022language,
  title={Language model-based emotion prediction methods for emotional speech synthesis systems},
  author={Yoon, Hyun-Wook and Kwon, Ohsung and Lee, Hoyeon and Yamamoto, Ryuichi and Song, Eunwoo and Kim, Jae-Min and Hwang, Min-Jae},
  journal={Proc. INTERSPEECH},
  year={2022}
}

@article{kalyanemotion,
  title={Emotion Arithmetic: Emotional Speech Synthesis via Weight Space Interpolation},
  author={Kalyan, Pavan and Rao, Preeti and Jyothi, Preethi and Bhattacharyya, Pushpak},
journal={Proc. INTERSPEECH},
year={2024}
}

@article{cho2024emosphere,
  title={EmoSphere-TTS: Emotional Style and Intensity Modeling via Spherical Emotion Vector for Controllable Emotional Text-to-Speech},
  author={Cho, Deok-Hyeon and Oh, Hyung-Seok and Kim, Seung-Bin and Lee, Sang-Hoon and Lee, Seong-Whan},
  journal={INTERSPEECH},
  year={2024}
}

@article{lotfian2017building,
  title={Building naturalistic emotionally balanced speech corpus by retrieving emotional speech from existing podcast recordings},
  author={Lotfian, Reza and Busso, Carlos},
  journal={IEEE Transactions on Affective Computing},
  volume={10},
  number={4},
  pages={471--483},
  year={2017},
  publisher={IEEE}
}

@article{lajszczak2024base,
  title={{BASE TTS}: Lessons from building a billion-parameter text-to-speech model on 100K hours of data},
  author={{\L}ajszczak, Mateusz and C{\'a}mbara, Guillermo and Li, Yang and Beyhan, Fatih and van Korlaar, Arent and Yang, Fan and Joly, Arnaud and Mart{\'\i}n-Cortinas, {\'A}lvaro and Abbas, Ammar and Michalski, Adam and others},
  journal={arXiv preprint arXiv:2402.08093},
  year={2024}
}

@article{kim2024clam,
  title={{CLaM-TTS}: Improving Neural Codec Language Model for Zero-Shot Text-to-Speech},
  author={Kim, Jaehyeon and Lee, Keon and Chung, Seungjun and Cho, Jaewoong},
  journal={Proc. International Conference on Learning Representations (ICLR)},
  year={2024}
}

@article{zhang2023speak,
  title={Speak foreign languages with your own voice: Cross-lingual neural codec language modeling},
  author={Zhang, Ziqiang and Zhou, Long and Wang, Chengyi and Chen, Sanyuan and Wu, Yu and Liu, Shujie and Chen, Zhuo and Liu, Yanqing and Wang, Huaming and Li, Jinyu and others},
  journal={arXiv preprint arXiv:2303.03926},
  year={2023}
}

@article{ekman1992argument,
  title={An argument for basic emotions},
  author={Ekman, Paul},
  journal={Cognition \& Emotion},
  year={1992},
  publisher={Taylor \& Francis}
}

@inproceedings{zhou2020converting,
  author={Kun Zhou and Berrak Sisman and Mingyang Zhang and Haizhou Li},
  title={{Converting Anyone’s Emotion: Towards Speaker-Independent Emotional Voice Conversion}},
  year=2020,
  booktitle={Proc. INTERSPEECH},
  pages={3416--3420}
}

@inproceedings{stanton2018predicting,
  title={Predicting expressive speaking style from text in end-to-end speech synthesis},
  author={Stanton, Daisy and Wang, Yuxuan and Skerry-Ryan, RJ},
  booktitle={2018 IEEE Spoken Language Technology Workshop (SLT)},
  pages={595--602},
  year={2018},
  organization={IEEE}
}

@article{russell1980circumplex,
  title={A circumplex model of affect.},
  author={Russell, James A},
  journal={Journal of Personality and Social Psychology},
  volume={39},
  number={6},
  pages={1161},
  year={1980},
  publisher={American Psychological Association}
}

@inproceedings{zhou2021seen,
  title={Seen and unseen emotional style transfer for voice conversion with a new emotional speech dataset},
  author={Zhou, Kun and Sisman, Berrak and Liu, Rui and Li, Haizhou},
  booktitle={Proc. IEEE International Conference on Acoustics, Speech and Signal Processing (ICASSP)},
  pages={920--924},
  year={2021}
}

@article{barra2010analysis,
  title={Analysis of statistical parametric and unit selection speech synthesis systems applied to emotional speech},
  author={Barra-Chicote, Roberto and Yamagishi, Junichi and King, Simon and Montero, Juan Manuel and Macias-Guarasa, Javier},
  journal={Speech Communication},
  volume={52},
  number={5},
  pages={394--404},
  year={2010},
  publisher={Elsevier}
}

@article{zhou2021limited,
  title={Limited Data Emotional Voice Conversion Leveraging Text-to-Speech: Two-stage Sequence-to-Sequence Training},
  author={Zhou, Kun and Sisman, Berrak and Li, Haizhou},
  journal={Proc. INTERSPEECH},
  year={2021}
}

@article{zhou2021emotional,
title = {Emotional voice conversion: Theory, databases and ESD},
journal = {Speech Communication},
volume = {137},
pages = {1-18},
year = {2022},
issn = {0167-6393},
author = {Kun Zhou and Berrak Sisman and Rui Liu and Haizhou Li}
}

@article{russell1977evidence,
  title={Evidence for a three-factor theory of emotions},
  author={Russell, James A and Mehrabian, Albert},
  journal={Journal of Research in Personality},
  volume={11},
  number={3},
  pages={273--294},
  year={1977},
  publisher={Elsevier}
}

@article{plutchik2001nature,
  title={The nature of emotions: Human emotions have deep evolutionary roots, a fact that may explain their complexity and provide tools for clinical practice},
  author={Plutchik, Robert},
  journal={American Scientist},
  volume={89},
  number={4},
  pages={344--350},
  year={2001},
  publisher={JSTOR}
}

@ARTICLE{zhou2022emotion,
  author={Zhou, Kun and Sisman, Berrak and Rana, Rajib and Schuller, Björn W. and Li, Haizhou},
  journal={IEEE Transactions on Affective Computing}, 
  title={Emotion Intensity and its Control for Emotional Voice Conversion}, 
  year={2023},
  volume={14},
  number={1},
  pages={31-48},
  doi={10.1109/TAFFC.2022.3175578}}

@inproceedings{parthasarathy2017jointly,
  title={Jointly Predicting Arousal, Valence and Dominance with Multi-Task Learning.},
  author={Parthasarathy, Srinivas and Busso, Carlos},
  booktitle={Proc. INTERSPEECH},
  pages={1103--1107},
  year={2017}
}

@article{wagner2023dawn,
  title={Dawn of the transformer era in speech emotion recognition: closing the valence gap},
  author={Wagner, Johannes and Triantafyllopoulos, Andreas and Wierstorf, Hagen and Schmitt, Maximilian and Burkhardt, Felix and Eyben, Florian and Schuller, Bj{\"o}rn W},
  journal={IEEE Transactions on Pattern Analysis and Machine Intelligence},
  volume={45},
  number={9},
  pages={10745--10759},
  year={2023},
  publisher={IEEE}
}

@inproceedings{zhou2024phonetic,
      title={Phonetic Enhanced Language Modeling for Text-to-Speech Synthesis}, 
      author={Kun Zhou and Shengkui Zhao and Yukun Ma and Chong Zhang and Hao Wang and Dianwen Ng and Chongjia Ni and Nguyen Trung Hieu and Jia Qi Yip and Bin Ma},
      year={2024},
      booktitle = {Proc. INTERSPEECH} 
}

@article{mcinnes2020,
  title={{UMAP}: Uniform Manifold Approximation and Projection},
  author={McInnes, Leland and Healy, John and Saul, Nathaniel and Gro{\ss}berger, Lukas},
  journal={Journal of Open Source Software},
  volume={3},
  number={29},
  year={2018}
}

@article{tang2023emomix,
  title={Emomix: Emotion mixing via diffusion models for emotional speech synthesis},
  author={Tang, Haobin and Zhang, Xulong and Wang, Jianzong and Cheng, Ning and Xiao, Jing},
  journal={INTERSPEECH},
  year={2023}
}

@ARTICLE{zhou2022speech,
  author={Zhou, Kun and Sisman, Berrak and Rana, Rajib and Schuller, Björn W. and Li, Haizhou},
  journal={IEEE Transactions on Affective Computing}, 
  title={Speech Synthesis With Mixed Emotions}, 
  year={2023},
  volume={14},
  number={4},
  pages={3120-3134},
  doi={10.1109/TAFFC.2022.3233324}}

@article{triantafyllopoulos2024expressivity,
  title={Expressivity and Speech Synthesis},
  author={Triantafyllopoulos, Andreas and Schuller, Bj{\"o}rn W},
  journal={arXiv preprint arXiv:2404.19363},
  year={2024}
}

@inproceedings{inoue2024hierarchical,
  title={Hierarchical Emotion Prediction and Control in Text-to-Speech Synthesis},
  author={Inoue, Sho and Zhou, Kun and Wang, Shuai and Li, Haizhou},
  booktitle={Proc. IEEE International Conference on Acoustics, Speech and Signal Processing (ICASSP)},
  year={2024}
}

@article{inoue2024fine,
  title={Fine-Grained Quantitative Emotion Editing for Speech Generation},
  author={Inoue, Sho and Zhou, Kun and Wang, Shuai and Li, Haizhou},
  journal={arXiv preprint arXiv:2403.02002},
  year={2024}
}

@article{wu2024laugh,
  title={Laugh Now Cry Later: Controlling Time-Varying Emotional States of Flow-Matching-Based Zero-Shot Text-to-Speech},
  author={Wu, Haibin and Wang, Xiaofei and Eskimez, Sefik Emre and Thakker, Manthan and Tompkins, Daniel and Tsai, Chung-Hsien and Li, Canrun and Xiao, Zhen and Zhao, Sheng and Li, Jinyu and others},
  journal={Proc. IEEE Spoken Language Technology Workshop (SLT)},
  year={2024}
}

@article{du2024cosyvoice,
  title={CosyVoice: A Scalable Multilingual Zero-shot Text-to-speech Synthesizer based on Supervised Semantic Tokens},
  author={Du, Zhihao and Chen, Qian and Zhang, Shiliang and Hu, Kai and Lu, Heng and Yang, Yexin and Hu, Hangrui and Zheng, Siqi and Gu, Yue and Ma, Ziyang and others},
  journal={arXiv preprint arXiv:2407.05407},
  year={2024}
}

@inproceedings{zhou2024learning,
  title={Learning Arousal-Valence Representation from Categorical Emotion Labels of Speech},
  author={Zhou, Enting and Zhang, You and Duan, Zhiyao},
  booktitle={Proc. IEEE International Conference on Acoustics, Speech and Signal Processing (ICASSP)},
  pages={12126--12130},
  year={2024}
}

\end{document}